\documentstyle[11pt,epsfig]{article}

\title{\bf Pomeron Vertices in Perturbative QCD in Diffractive
Scattering}
\author{J.Bartels$^{a}$, M.Braun$^{a,b}$ and G.P.Vacca$^{a,c}$ \\
$^a$ II Institut f\"ur Theoretische Physik, Univ. of Hamburg, Germany\\
$^b$Department of High Energy Physics, Univ. of S.Petersburg,
Russia\\
$^c$Dipartimento di Fisica and INFN, Univ. of Bologna, Italy }
\def\beq{\begin{equation}}
\def\eeq{\end{equation}}
\def\bea{\begin{eqnarray}}
\def\eea{\end{eqnarray}}

\def\o{\omega}
\begin{document}
\maketitle
\medskip
\noindent
{\bf Abstract:}
We analyse the momentum space triple Pomeron vertex in perturbative QCD.
In addition to the standard form of this vertex which is used in
the context of total cross-sections at high energies and
in the QCD reggeon field theory,
there exists an alternative form which has to be used in the study
of high-mass diffraction.
We review and analyse the relation between these two versions.
We discuss some implications for the BK-equation.
In the second part of our paper we extend this
analysis to the Pomeron-Odderon-Odderon vertex.

\section{Introduction}
It is well known that the BFKL Pomeron - which in perturbative QCD
describes the high energy elastic scattering of small-size projectiles -
is part of an effective 2+1 dimensional field theory. Apart from the
BFKL kernel (and its generalization to the kernel of the BKP equation),
the most essential element of this field theory, known so far, is the vertex
which describes the transition of two to four reggeized gluons;
as the most important application, it leads to the splitting of one BFKL
Pomeron into two Pomerons (PPP vertex).
This vertex was first derived in the study of high-mass diffractive
scattering of two virtual photons
\cite{JB,JBMW}.
The triple Pomeron vertex appears after summing 
and integrating over the diffractive final states and removing 
a certain piece which can be associated with a single Pomeron
exchange (see some details below).
In the context of the 2+1 dimensional field theory, this vertex
can then be used to compute, for example, the Pomeron self-energy.
Another application is the evolution equation for the sum of
the fan diagrams
which, in the large-$N_c$ limit, coincides with the
Balitsky-Kovchegov (BK)
equation ~\cite{BK,MAB1,BLV2}. Also, this vertex easily allows to
understand the AGK counting rules in pQCD.
On the other hand, in the Regge-Gribov formalism high-mass diffraction
is standardly ascribed exclusively to the triple Pomeron interaction.
This suggests that there may exist an alternative 
form of the triple Pomeron vertex, the diffractive PPP vertex,
suitable for the description of high-mass diffraction.

To illustrate this difference it is useful to recapitulate briefly the
derivation of the former PPP-vertex.
Ref.\cite{JBMW} introduces amplitudes which describe the
coupling of virtual photons to four $t$-channel gluons,
named reggeon-particle amplitudes, $D_4$. They are
derived from multiple energy discontinuities and include the integration
over the rhs cuts of the energy variables. One starts from
gluon production amplitudes with exchanges of reggeized gluons,
and then sums over the number of produced gluons and integrates over
the produced mass. In the language of angular
momentum theory, however, this function $D_4$ is not yet a $t$-channel
partial wave,
since the lhs energy-cuts have not yet been included, and signature
has not been defined. As outlined in \cite{JBMW}, the final step for
obtaining signatured partial waves is the decomposition of $D_4$ into pieces
with definite symmetries with respect to the interchange of reggeized
$t$-channel gluons. In particular, it was observed that there are
pieces of $D_4$ which are
antisymmetric under the exchange of two gluons and satisfy bootstrap equations.
As a result of these bootstrap equations, these pieces of $D_4$ can be
reduced to BFKL amplitudes, $D_2$, with two reggeized gluons only.
The sum of all these reggeizing pieces has been named $D_4^R$
(the superscript 'R' refers to `reduction').
After subtracting $D_4^R$ from $D_4$,
one is left with a new amplitude, $D_4^I$, which is totally symmetric
under the interchange of the outgoing gluons:
\begin{equation}
\label{decompD4}
D_4 (1,2,3,4) = D_4^R (1,2,3,4) + D_4^I (1,2,3,4)\,.
\label{decomp}
\end{equation}
This amplitude $D_4^I$ now has all the properties of the partial wave,
and it contains the $2 \to 4$ gluon vertex $V$ which can be used to
define the triple Pomeron vertex. This vertex is one of the fundamental
interaction terms in the BFKL field theory. As outlined in
~\cite{GPT,ABSW}, reggeon field theory is, by construction, a
solution to the set of $t$-channel reggeon unitarity equations, and these
equations require the $t$-channel partial waves to have the correct
symmetry and partial wave properties.

The decomposition (\ref{decomp}), in a diagrammatic fashion,
is illustrated in Fig.~1.
A striking feature is the absence of the double Pomeron exchange
(Fig.~2) on the rhs of eq.(\ref{decomp}),
which has been included in the definition of $D_4$ on the lhs. 
At first sight, the disappearance
of this contribution appears to be somewhat
strange. A closer look at the discussion presented in ~\cite{JBMW}, in
particular at the form of the vertex $V$, shows that most of the double
Pomeron exchange has been absorbed into the 3P
interaction part $D_4^{I}$: when analysing the momentum structure
of the vertex $V$, one finds pieces that do not correspond to an $s$-channel
intermediate state but rather to a virtual correction. This is quite analogous
to the LO BFKL kernel which consists of a `real' piece, corresponding to the
production of an $s$-channel gluon, and a virtual piece, coming for the
gluon trajectory. As consequence of these virtual corrections inside
the vertex $V$, $D_4^I$ contains most of the double Pomeron exchange;
some remainders have also been absorbed into $D_4^R$. At very high rapidities,
$D_4^I$ dominates, since it involves two Pomerons
coupled to the target: the decomposition (1), therefore, also allows to
separate the leading term of the amplitude $D_4$ in the high-energy limit.

\begin{figure}
\begin{center}
\epsfig{file=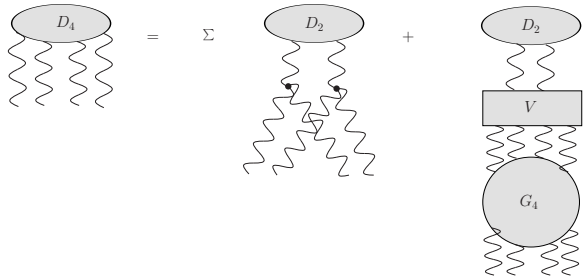,width=8cm}
\caption{ Decomposition for $D_4=D_4^R+D_4^I$.}
\end{center}
\label{figD4}
\end{figure}

\begin{figure}
\begin{center}
\epsfig{file=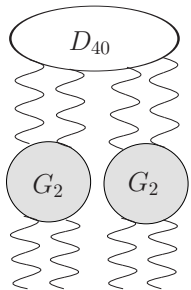,width=3cm}
\caption{ Double Pomeron exchange
($G_2$ is the BFKL Pomeron Green's function).}
\end{center}
\label{figDPE}
\end{figure}

Nevertheless, the decomposition (1) has a certain drawback.
If one studies high-mass diffraction in pQCD, multigluon intermediate states
are contained in both pieces in (1), i.e. this decomposition
does not clearly separate the diffractive production of $q\bar{q}$,
$q\bar{q}g$, $q\bar{q}gg$... final states.
As mentioned, we are used to write the diffractive cross section
formula
in terms of a triple Pomeron coupling. Thus one might suspect that
there is another, diffractive triple Pomeron vertex, which
completely describes high-mass diffraction
 but perhaps is
less convenient for the study of the high-energy limit.
In fact, as was already noticed in ~\cite{JBMW}, the definition
of the 3P vertex is not unique. A different form of the 3P vertex
was later proposed in ~\cite{MAB2} and analysed in ~\cite{MBGV},
which precisely corresponds to the definition of such a `diffractive
vertex'. The idea was,
starting from the same $D_4$ and using the large $N_c$ limit,
to find an alternative decomposition, in which,
instead of the reggeizing term $D_4^{R}$, the double Pomeron exchange
is separated. It turns out that the remainder then, again,
contains a triple Pomeron interaction, which was named $Z$ and differs from
the previous version, $V$.
In contrast to (1), such a decomposition no longer picks out the
leading part of the amplitude in the high-energy limit,
since the double P exchange and 3P part behave similarly.
However, for the description of exclusive high-mass
diffractive processes
such a decomposition is much more natural:
the first term contains only the
$q\bar{q}$, but no multigluon intermediate states, whereas the second
term has all the diffractive multigluon intermediate states,
quite in accordance with the physical picture.

This diffractive vertex $Z$ was first derived in
~\cite{JBMW}, in the direct calculation of the high-mass diffractive
cross-section at $t=0$. In the lowest order of $\alpha_s$, the
cross-section for $q\bar{q}g$ production, integrated over the
quark and gluon momenta, can be written in terms of the
diffractive vertex
$Z$ which couple to elementary $t$-channel gluons.
So the 3P vertex $Z$ contains only real gluon production ( and no
virtual corrections)\footnote{This statement does not preclude that
certain pieces of $Z$ can be expressed in terms of the gluon trajectory
function: this feature is will known from the BFKL bootstrap equation,
where parts of the real gluon production inside the BFKL kernel
can be expressed in terms of the gluon trajectory function}.
This is in contrast to the triple Pomeron vertex $V$ which has been discussed
above.

All these considerations remain valid also for more compliciated cases,
in particular for diffractive processes governed by the exchange of
Odderons. The simplest case, the diffractive production of $q\bar{q}$
states (with suitable quantum numbers), can be described by the exchange
of two Odderons which in pQCD are bound states of three
reggeized gluons. Then comes the production of $q\bar{q}g$,
$q\bar{q}gg$,... states, and, as shown in ~\cite{BE},
after summation over the number of gluons and integration over
their momenta one arrives at an amplitude $D_6$ with 6 reggeized
$t$-channel
gluons and at a decomposition, analogous to (1). As a part of this
decomposition, a vertex for the transition $2 \to 6$ gluons, the
Pomeron$\to$Odderon-Odderon (POO) vertex was found ~\cite{BE}.
Here, again, the question arises whether a `diffractive' version
of this vertex exists and what form it has.

In the present paper, after clarifying first the relation
between the vertices $V$ and $Z$ for the triple Pomeron case,
we generalize the argument to the transition $P\to OO$ and derive the
diffractive POO vertex $Z$. Following the derivation of $Z$ in
~\cite{JBMW}, one can derive this vertex by studying,
in the lowest order, the integrated gluon production cross section
of the process $\gamma^*q \to (q\bar{q}g)q$ where the outgoing diffractive
state has $C$-parity opposite to the incoming photon.
In this paper, however, we propose a simpler method which,
in the large $N_c$ limit, allows to avoid these calculations and
readily leads to the desired diffractive vertex.

This paper is organized as follows.
In section 2 we review the discussion of the PPP
vertex, and we re-derive the explicit form of the `diffractive' PPP vertex.
In section 3 we turn to the Odderon case and obtain the diffractive
POO vertex. A few details are put into 2 appendices.
For simplicity, throughout the paper we will restrict ourselves
to the large-$N_c$ limit.
\section{ Diffractive vertex for the splitting P$\to$ PP}
\subsection{Definitions}
The diffractive vertex is derived from the high energy limit of the
diffractive cross section in perturbative QCD ~\cite{JB,JBMW,MAB2}.
To be definite, one considers the process $\gamma^* +q \to (q\bar{q}+n\;g) +q$
in the triple Regge limit $Q^2 \ll M^2 \ll s$, where $Q^2$, $M^2$ and $s$
denote the squared mass of the virtual photon, the squared mass of the
diffractively produced system, and the square of the total energy, resp.
The cross section formula for this process defines amplitudes with four
reggeized gluons in the $t$-channel, $D_4$.
To explain our notation, we remind of the BFKL amplitude, $D_2$,
which appears in the elastic scattering process $\gamma^* +q \to \gamma^* +q$:
it describes the exchange of two reggeized gluons, and it satisfies the
BFKL equation~\cite{BFKL} which we write in the operator form
\beq
\label{BFKL}
S_{2}D_{2}=D_{20}+V_{12}\otimes D_{2}\,,
\eeq
where
$ S_{2}=j-1-\omega(1)-\omega(2) $ is the two-gluon  "free"
Schr\"odinger operator for the energy. Here
$ \omega(k) $ is the gluon Regge trajectory function, and
$ V_{12} $ is the BFKL interaction kernel (containing only the real gluon
production). Here the subscripts `1', `2' refer to gluon `1', `2' with momenta
$k_1$, $k_2$, and color $a_1$, $a_2$.
We suppress momentum variables and color indices. Powers of the
coupling $g^2$ and color tensors are contained in $V_{12}$ and will not be
written explicitly.
Furthermore, we simplify by writing
\beq
D_{20}(1)\equiv D_{20}(1,2) \,,
\eeq
since our discussion will be done for fixed  momentum
transfer $r^2$, and the variables for the second gluon follow from
momentum conservation $k_2 = r-k_1$ etc. 

\begin{figure}
\begin{center}
\epsfig{file=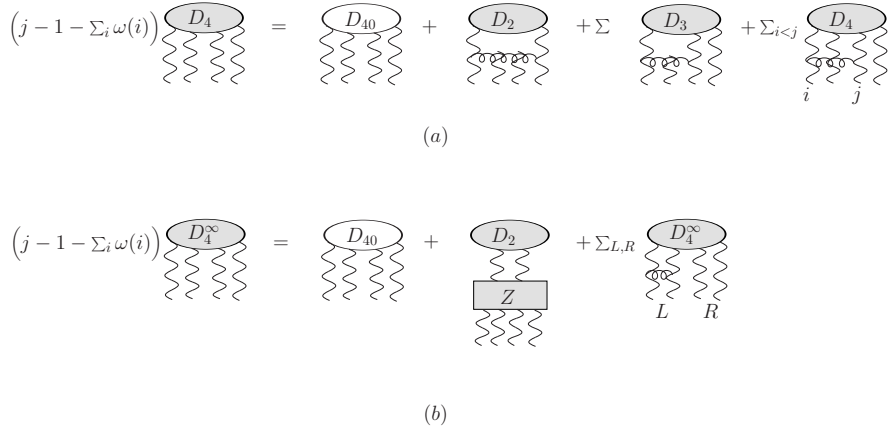,width=12cm}
\caption{(a) Graphical illustration of the definition of $D_4$; 
(b) the same equation after color projection and large $N_c$ limit.}
\end{center}
\label{figD4eq}
\end{figure}

To define the 'diffractive triple Pomeron vertex' we start from the
integral equation for $D_4$ ~\cite{JBMW} which sums the diagrams for the
diffractive production of a $q\bar{q}$ pair plus $n$ gluons.
The equation is illustrated
in Fig.~$3$a: we take the total transverse momentum of all
four gluons equal to zero, but we keep in mind that each
of the gluon pairs may have a nonzero total momentum $q$ and $-q$, with
$t=-q^2$ being the invariant momentum transfer squared in the
diffraction process. Each vertex corresponds to real gluon production, and all
$t$-channel gluon lines denote reggeized gluons. As a result, one readily
understands the content of this equation in terms of $s$-channel
intermediate states. Next, using the bootstrap properties of $D_3$
the second and third group of terms can be combined to an effective vertex,
$\tilde{V}$:
\beq
S_4 D_4=D_{40}+\tilde{V} \otimes D_2+U\otimes D_4,
\eeq
where $S_4$ is defined in analogy with $S_2$ (after Eq.(\ref{BFKL})),
and $U$ denotes the interactions between the lower gluons:
\beq
U=V_{12}+V_{34}+V_{13}+V_{14}+V_{23}+V_{24}\,.
\eeq
Since, in our diffractive cross section, at the lower end the two pairs of
gluons (1,2) and (3,4) have to be in color singlets,
we introduce a suitable color projector $P$ which acts on the color indices
of the lower gluons. In the limit of large $N_c$ we use
\beq
P U \otimes D_4 \approx P U_0 \otimes P D_4,
\label{PPproj}
\eeq
where
\beq
U_0 = V_{12}+V_{34}
\eeq
denotes the rungs inside the pair (1,2) and (3,4).
Introducing the notation $D_4^{\infty} = \lim_{N_c \to \infty} P D_4$
the integral equation becomes (Fig.~$3$b):
\beq
S_4 D_4^{\infty} = PD_{40} + Z \otimes D_2 + PU_0 \otimes D_4^{\infty}.
\label{Zdef}
\eeq
The equation has been derived in ~\cite{MAB2}.
$Z=\lim_{N_c \to \infty} P \tilde{V}$ stands for the 'diffractive triple
Pomeron vertex':
it describes the coupling of the two lower gluon ladders to the diffractive
system. In an operator notation, it acts on the forward
BFKL function $D_2$ and transforms the $t$-channel two gluon state
into the diffractive colour state of four gluons. Note the explicit
appearance of the double Pomeron exchange in the inhomogeneous term:
when solving the integral equation by iteration, we generate the
two Pomeron ladders (Fig.2) for the diffractive $q\bar{q}$ intermediate state.

Let us now discuss the connection of this vertex $Z$ with the triple Pomeron
vertex $V$ obtained in ~\cite{JBMW}.
This analysis will help us to find the explicit form of the diffractive
vertex $Z$, and, later on, the same method can be applied to the
Odderon case. Starting from the result of ~\cite{JBMW}
\beq
D_4=D_4^R + D_4^I\,,
\label{D4decomp}
\eeq
we remind that the second part, $D_4^I$, satisfies the equation (Fig.~4)
\beq
\label{D4Idef}
S_4 D_4^I = V \otimes D_2 + U \otimes D_4^I\,,
\eeq
where $V$ is the $2 \to 4$ gluon vertex. It consists of three pieces
and it can be expressed in terms of one single function:
\beq
\label{Vdef}
V = V(1,2,3,4) + V(1,3,2,4) + V(1,4,2,3)\,.
\eeq

\begin{figure}
\begin{center}
\epsfig{file=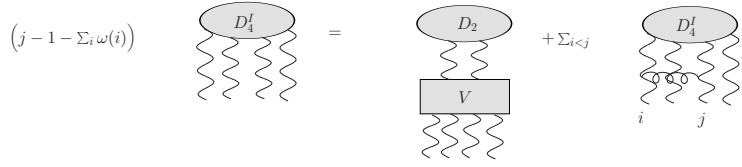,width=10cm}
\caption{ Defining equation for $D_4^I$.}
\end{center}
\label{figD4Ieq}
\end{figure}

When projecting on color singlet states in the pairs (1,2) and (3,4),
in the limit $N_c \to \infty$ the first (planar) term represents the
leading part, the remaining two terms are nonplanar and subleading
in $1/N_c$.  Note that this vertex $V$ can be obtained from
$D_4$ only after the removal of the reggeizing pieces $D_4^R$.
As a result of this `reduction' procedure, the vertex $V$ contains
both `real ' and `virtual' contributions, very much in the same way as
the full BFKL kernel has both real and virtual pieces:
the latter ones belong to the reggeization of the gluon.
In this sense the connection between $V$ and real gluon production is
more complicated than in the case of $Z$.

Now let us rewrite the equation (\ref{D4Idef}). Insert $D_4^I = D_4 - D_4^R$
and regroup the terms:
\beq
S_4 D_4 = S_4 D_4^R - U \otimes D_4^R + V \otimes D_2 + U \otimes D_4\,.
\label{D4interm}
\eeq
Making use of the color projector $P$ and taking the limit $N_c \to \infty$
(with help of the identity (\ref{PPproj})) we find:
\beq
S_4 D_4^{\infty} = (P S_4 - P U_0 \otimes) D_4^R + PV \otimes D_2 +
P U_0 \otimes D_4^{\infty}\,.
\label{D4interminfty}
\eeq
Here the large-$N_c$ limit of $PV$ selects the planar part of the
$2 \to 4$ gluon vertex. Comparison of (\ref{D4interminfty}) with
(\ref{Zdef}) leads to
\beq
Z\otimes D_2 = P \left(V\otimes D_2 +(S_4-U_0 \otimes) D_4^R - D_{40}
\right)\,.
\label{VZ-connection}
\eeq
This equation establishes the relation between the two versions of the
triple Pomeron vertex. The explicit expression for $Z$ was first obtained
in ~\cite{MAB2} and its relation with V derived in~\cite{MBGV,vacca_phd}.

The important point is that the original form (\ref{Zdef}) has a very
transparent physical interpretation.
The solution of (\ref{Zdef}) is a sum of two terms, each with two pomeron legs
attached to them. The first, coming from $D_{40}$, has the meaning of double
pomeron exchange with only $q\bar q$ intermediate states (low mass
diffraction). The second, involving $Z$, corresponds to high-mass
diffraction, with many gluons in the intermediate states. So one would
expect that the inclusive cross-section will be given in terms of this vertex
$Z$ coupled to two pomerons at rapidity $y$. This expectation is confirmed
~\cite{JBMW} by the calculation of the diffractive cross-sections at $t=0$
in the lowest (non-trivial) order, i.e. of the cross section for
diffractive $q\bar{q}$ production. Below we will show that
the expression for $Z$ coincides with the diffractive vertex in ~\cite{JBMW}.

Finally note that in all our discussion we have kept rapidity $y$
fixed. If one is
interested in cross-sections integrated over the rapidities of
observed jets, it is more convenient to use the total amplitude in the form
(\ref{D4decomp}), i.e. to write the the cross-section as a sum of the
reggeized zero-order term directly coupled to the initial and final
projectiles plus the contribution from the triple pomeron coupling
with the vertex $V$.
\subsection{The explicit form of the diffractive triple Pomeron vertex}
In this subsection we use Eq.~(\ref{VZ-connection}) and derive an explicit
expression for the diffractive vertex $Z$.
As found in ~\cite{JB,JBMW}, the explicit form of $D_4^R$ is given by
\beq
D_4^R=C\Big\{\sum_{i=1}^4D_2(i)-D_2(12)-D_2(13)-D_2(14)
\Big\}.
\label{D4R}
\eeq
(here the argument `12' is a shorthand notation for the sum
of the momenta $k_1+k_2$).
For the momenta we will assume that $k_1 + k_2 = -k_3 -k_4=q$;
in short: $12=-34=q$. The coefficient $C=(1/2)g^2$ will be suppressed in the
following for brevity and has to be re-added in front of our final expressions.
In the following (and in a similar derivation in Sec. 3,
Eqs. (\ref{X6ini})-(\ref{X6final})) we assume that the colour projector $P$
has been applied to operators acting on final gluons,
which we do not specify explicitly.

At its lower end the vertex $Z\otimes D_2$ couples to two color singlet
pairs, each being symmetric in its transverse momenta, and the whole
expression has to be symmetric under the interchange of the two pomerons.
So the quantity we have to study,
\beq
X=\Bigl((j-1)-\sum_{i=1}^4\o(i)-U_0\Bigl)D_4^R-D_{40}\,,
\label{Xexpr}
\eeq
is to be convoluted with a function which is symmetric in the momenta (12)
and (34), and also under the interchange $(12)\leftrightarrow (34)$.
Since the operator $S_4-PU_0$ also possesses this symmetry,
it follows that  for our purpose it would be sufficient to take
\beq
D_4^R=4D_2(1)-D_2(12)-2D_2(13).
\label{D4Rred}
\eeq
Nevertheless, we shall consider the general form valid for the a
generic momentum configuration $(1,2,3,4)$ of the four $t$-channel gluons,
and we impose the restriction to the diffractive case
$12=-34=q$ only at the end of our calculation.

Now we evaluate eq. (\ref{Xexpr}). The simplest term is the first one: 
we have to apply the operator $(S_4-U_0)$ to the
expression (\ref{D4R}). 
We write the nonforward inhomogeneous
BFKL equation (\ref{BFKL}) in a slightly different form:
\begin{eqnarray}
(j-1) D_2(i) & = & D_{20}(i) + (V_{12} + \omega(1) + \omega(2)) \otimes D_2
\nonumber \\
& =& (j-1)D_2(i)=D_{20}(i)+G(i,0,1234-i)\, ,
\end{eqnarray}
where $G(1,2,3)$ is the infrared finite and conformally invariant
function introduced in ~\cite{JBMW,MBGV,vacca_phd}. It has the form
\beq
G(1,2,3)=G(3,2,1)= - W(1,2,3)-D_2(1)[\omega(2)-\omega(23)]
-D_2(3)[\omega(2)-\omega(12)]\,,
\eeq
where
\beq
W(1,2,3)=\int\frac{d^2k'_1}{(2\pi)^3}K_{2\to 3}(1,2,3|1',3')D_2(1')\,,
\label{Wdef}
\eeq
\beq
K_{2\to 3}(1,2,3;1',3')=V_{13}(2,3;1-1',3')-V_{12}(12,3;1',3')\,,
\eeq
and on the rhs of the last equation we have written explicitly the incoming
and outgoing momenta of the BFKL kernel $V_{ij}$.
Taking into account (\ref{D4R}) we find
\beq
(j-1)D_4^R=D_{40}+\sum_{i=1}^{4} G(i,0,1234-i)
-G(12,0,34)-G(13,0,24)-G(14,0,23)\,.
\eeq
Continuing with the rhs of Eq.~(\ref{Xexpr}), we postpone the terms
$-\sum_{i=1}^4\o(i) D_4^R$: later on we shall see that they will cancel
with similar terms coming from the other parts of the operator. So we have
to study only the terms with $U_0$ wihich we will denote by
\beq
\tilde{X}=-U_0 \otimes  D_4^R\,.
\eeq
The final $X$ will be given by
\beq
X=\sum_{i=1}^{4} G(i,0,1234-i)
-G(12,0,34)-G(13,0,24)-G(14,0,23)-\sum_{i=1}^4\o(i)D_4^R+\tilde{X}\,.
\label{Xparz}
\eeq

For the further investigation of $X$ we shall make use of two
identities derived earlier:\\
1) the bootstrap relation~\cite{LNL}:
\beq
V_{23}D_2(1)=2[\omega(23)-\omega(2)-\omega(3)]D_2(1)\,;
\label{bootrel}
\eeq
2) the action of an interaction on $D_2$ with a shifted argument~\cite{MAB2}:
\beq
V_{12}D_2(13)=W(24,1,3)+W(4,2,13)-W(3,12,4)-W(13,0,24)\,.
\label{shiftrel}
\eeq
Here the function $W$ has been defined in (\ref{Wdef}).
A special case of (\ref{Wdef}) is
\beq
V_{13}(1,3;1',3')\otimes D_2(1)=-W(1,0,3)\,.
\eeq

Armed with these identities we may start calculating various terms in $X$,
similarly to {\cite{MBGV,vacca_phd}. We begin with
\beq
\tilde{X}_1=-U_0 \otimes D_2(1)=- (V_{12}+V_{34})\otimes D_2(1)\,.
\eeq
The term independent of 1 will give :
\beq
- V_{34}\otimes D_2(1)= -2[\omega(34)-\omega(3)-\o(4)]D_2(1)\,.
\eeq
The term with $V_{12}$ leads to
\beq
V_{12}\otimes D_2(1)=W(1,2,34)-W(1,0,234)\,.
\eeq
Summing the two terms leads to
\beq
\tilde{X}_1=[W(1,0,234)-W(1,2,34)]
-2[\omega(34)-\omega(3)-\omega(4)]D_2(1)\,.
\eeq
An analogous result with just a simple permutation of the momentum
labels is
easily obtained for the terms $-U_0 \otimes D_2(2)$, $-U_0 \otimes D_2(3)$,
and $-U_0 \otimes D_2(4)$.
Now we pass to the second term in (\ref{D4Rred}) and compute
\beq
\tilde{X}_{12}=-U_0 \otimes D_2(12)=-(V_{12}+V_{34})\otimes D_2(12)\,.
\eeq
Here, due to the bootstrap relation:
\beq
\tilde{X}_{12}=
-2[\omega(12)-\omega(1)-\omega(2)+\omega(34)-\o(3)-\omega(4)] D_2(12)\,,
\eeq
both terms can be expressed in terms of trajectory functions.
From the third term in (\ref{D4Rred}) we have
\beq
\tilde{X}_{13}=-U_0 \otimes D_2(13) = - (V_{12}+V_{34})\otimes D_2(13)\,.
\eeq
Here both terms are non-trivial:
\beq
V_{12}\otimes D_2(13)=W(24,1,3)+W(4,2,13)-W(4,12,3)-W(24,0,13)\,,
\eeq
\beq
V_{34}\otimes D_2(13)=W(13,4,2)+W(1,3,24)-W(1,34,2)-W(24,0,13)\,.
\eeq
In the sum we obtain:
\[
\tilde{X}_{13}=[ 2W(13,0,24)+W(1,34,2)+W(3,12,4)\]
\beq
-W(1,34,2)-W(2,4,13)
-W(3,1,24)-W(4,2,13)]\,.
\eeq

In our next step we rewrite these expressions in terms of the $G$ function.
This is lengthy but straightforward. By definition
\beq
G(1,2,3)=G(3,2,1)=-W(1,2,3)-D_2(1)[\omega(2)-\omega(23)]
-D_2(3)[\omega(2)-\omega(12)]\,,
\eeq
from which we arrive at:
\beq
W(1,2,3)=-G(1,2,3)-D_2(1)[\omega(2)-\omega(23)]
-D_2(3)[\omega(2)-\omega(12)]\,.
\label{WGrel}
\eeq
So all we have to do is to substitute this expression for $W$
into our formulas for the different $\tilde{X}$.
This substitution obviously expresses $W$ through $G$ functions
and through terms with $\omega$'s. These
additional $\omega$ terms can be combined into a simple
combination (Appendix 1) which, when combined with
similar terms in eq. (\ref{Xparz}), cancels.

As a result, summing all terms as indicated in (16) we get
\bea
\tilde{X}\!\!&=&\!\!\!\!\!\!-U_0 \otimes  D_4^R=
G(1,2,34)+G(2,1,34)+G(12,3,4)+G(12,4,3)+2 G(1,34,2)+2 G(3,12,4)
\nonumber \\
&&-G(1,4,23)-G(4,1,23)-G(14,2,3)-G(14,3,2) -G(1,3,24)-G(3,1,24)
\nonumber \\
&&-G(2,4,13)-G(4,2,13)+2 G(13,0,24)+2 G(14,0,23)\nonumber \\
&&-G(1,0,234)-G(2,0,341)-G(3,0,412)-G(4,0,123)+\sum_{i=1}^4\o(i)D_4^R\,.
\label{Xtildeexpr}
\eea
Inserting this into (\ref{Xparz}), the terms proportional to
$\sum_{i=1}^4\o(i)D_4^R$ cancel, and $X$ can be expressed entirely in terms
of $G$ functions:
\bea
X\!\!&=&
G(1,2,34)+G(2,1,34)+G(12,3,4)+G(12,4,3)+2 G(1,34,2)+2 G(3,12,4)
\nonumber \\
&&-G(1,4,23)-G(4,1,23)-G(14,2,3)-G(14,3,2) -G(1,3,24)-G(3,1,24)
\nonumber \\
&&-G(2,4,13)-G(4,2,13)+G(13,0,24)+G(14,0,23)-G(12,0,34) \, .
\label{XGexpr}
\eea

Returning to the lhs of Eq.~(\ref{VZ-connection}), we express $V$
in terms of $G$'s ~\cite{JBMW}:
\bea
 V(1,2,3,4) = \lim_{N_c \to \infty} P V\otimes D_2 & = &
G(1,23,4)+G(2,13,4)+G(1,24,3)+G(2,14,3) -G(12,3,4)
\nonumber\\
&&-G(12,4,3)-G(1,2,34)-G(2,1,34)+G(12,0,34)\,.
\eea
Using Eq.(\ref{VZ-connection}) for $Z$ and Eq. (\ref{XGexpr})
leads to our final expression:
\bea
Z(1,2,3,4) &\equiv& Z\otimes D_2=
G(1,23,4)+G(2,14,3)+G(1,24,3)+G(2,13,4)\nonumber \\
&&-G(13,2,4)-G(13,4,2)-G(1,3,24)-G(3,1,24)\nonumber\\
&&-G(14,2,3)-G(14,3,2)-G(1,4,23)-G(4,1,23)\nonumber\\
&&+2G(1,34,2)+2G(3,12,4)+G(13,0,34)+G(14,0,23)
\,,
\eea
which is the result firstly derived in \cite{MBGV,vacca_phd},
where it was already observed that $Z$ can also
be expressed in terms of the nonplanar pieces of the $V$ function
 (cf. Eq.~(\ref{Vdef}) in ~\cite{JBMW})
\beq
\label{Zexplicit}
Z(1,2,3,4)=V(1,3,2,4)+V(1,4,2,3)\,.
\eeq

Let us finally compare our result with the diffractive vertex derived in
~\cite{JBMW}.
We consider diffractive kinematics, i.e. we put the total transverse
momentum equal to zero and we assign the momenta:
\[
1=l,\ \ 2=-l,\ \ 3=m,\ \ 4=-m,\ \ (t=0)\,.
\]
In this configuration we find, using the notation of ~\cite{JBMW} for the
$G$ function which depends only on two arguments:
\bea
Z \otimes D_2 \!\!\!&=& \!\!\!2G(l,-l)+2G(m,-m)+G(l,-m)+2G(l,m)+G(l+m,-l,-m)
+G(l-m,m-l)\nonumber \\
&&\!\!\!-2G(l,m-l)-2G(-l,m+l)-2G(m,l-m)-2G(-m,l+m) \,.
\eea
This coincides with the expression for the diffractive cross-section
at $t=0$ in ~\cite{JBMW}; in contrast to the triple Pomeron
vertex $V$ which also contains virtual contribtions this diffractive
triple Pomeron contains only the real contribution for produced gluons.
Note that the diffractive vertex in ~\cite{JBMW} has been derived
without any large-$N_c$ approximation.

Let us comment on our result. Equation (\ref{Zdef}) (Fig.~3b) which defines the
diffractive vertex $Z$ contains, as an inhomogeneous term, the coupling
of two color singlet pairs of gluons to the $q\bar{q}$ system: solving
the equation by iteration generates the two Pomeron ladders which
are needed to derive a cross section formula for the diffractive production
of a $q\bar{q}$ pair (double Pomeron exchange).
This to be compared with equation (\ref{D4Idef}) (Fig.~4) for
$D_4^I$ where such a term is absent; also $D_4^R$, which is a sum of
BFKL ladders $D_2$, does not have such a term. On the other hand, the
function $D_4$ from which both $D_4^I$ and $D_4^R$ are obtained contains this
contribution. The explanation is the following: while splitting $D_4$ into
the sum $D_4^R + D_4^I$ we regroup the QCD diagrams in a special way;
as a consequence of the bootstrap, the two-Pomeron ladder
diagrams are absorbed into $D_4^R$ and into $V \otimes D_{20}$. In the latter
piece, the double Pomeron exchange becomes part of the virtual contributions
inside the triple Pomeron vertex $V$.

A second comment applies to the transformation to configuration space.
Whereas the momentum space representation of high energy QCD allows
a direct control of the $s$-channel unitarity content of scattering
amplitudes, in recent years an appealing intuitive picture of high energy
scattering processes has been developed in configuration space,
the QCD dipole picture~\cite{dipole}. A popular model is the nonlinear
Balitsky-Kovchegov (BK) equation ~\cite{BK}, whose structure is reminiscent
of the fan diagram equation of ~\cite{GLR}. In a recent paper ~\cite{BLV2}
the BK equation has been compared with the nonlinear equation which sums
fan diagrams built from BFKL ladders. Making use of symmetry properties
of the BFKL kernel, it has been shown, for a particular set of initial
conditions, that in the large-$N_c$ limit the kernel of the BK-equation
coincides with the Fourier transform of the triple Pomeron vertex $V$.
From this one concludes that the BK equation 'knows' about the
two-Pomeron ladder diagrams, but it contains only a part of them.

Conversely, one might want to know the Fourier transform of the diffractive
triple Pomeron vertex $Z$, eq.(\ref{Zexplicit}). As shown
in ~\cite{BLV2} the first nonleading (in 1/$N_c^2$) corrections to the
fan diagram equation are due to the nonplanar part of the triple Pomeron
vertex $V$ which coincides with the rhs of Eq.~(\ref{Zexplicit}),
$V(1,3,2,4)+V(1,4,2,3)$. Its Fourier transform turns out to be simple,
but quite different from the BK kernel. This, once more, confirms that the BK
kernel contains a part of the two Pomeron ladder diagrams: the diffractive
vertex $Z$, by definition, excludes these contribtions, and its Fourier
transform is different from the BK kernel. As a conclusion, a full account
of the $q\bar{q}$ diffractive intermediate state cannot be obtained by
simply adding this contribution to the BK equation; this would lead to an
overcounting of QCD diagrams.
\section{ Diffractive vertex for the splitting P$\to$ OO}
\subsection{Definitions}
We now turn to six gluons and repeat the same line of arguments.
We start from the integral equation for $D_6$ ~\cite{BE} which 
sums all gluon production diagrams (Fig.~5). It has the structure
\beq
S_6 D_6 = D_{60} + \sum K_{2 \to 6}\otimes D_2 +\sum K_{2 \to 5}\otimes D_3
+ \sum K_{2 \to 4}\otimes D_4 +\sum K_{2 \to 3}\otimes D_5 + U \otimes D_6\,,
\label{D6inteq}
\eeq
where 
\beq
U = \sum_{(i,j)} V_{ij}
\eeq
denotes the sum over all pairwise interactions, and 
$S_{6}=j-1-\sum_{i=1}^6\omega(i)$ is the natural extension of $S_4$ 
for the 6 reggeized gluon propagator. Invoking the bootstrap 
properties of $D_3$ and $D_5$ and the decomposition (\ref{D4decomp}) 
for $D_4$, one immediately sees that, on the rhs of (\ref{D6inteq}),
the second and the third term, the $D_4^R$ pieces in the fourth and the 
fifth term all start with a two-gluon states below the quark loop and 
then define an effective $2 \to 6$ vertex, $\tilde{Z}$. 
Let $P$ be a projector on the 
two Odderon state (Odderons are formed by the triplets (123) and (456)), 
and let $P$ act on both sides of (\ref{D6inteq}).
Next we take the large $N_c$ limit: define 
$D_6^{\infty}=\lim_{N_c\to\infty} P D_6$ and 
\beq 
U_0=V_{12}+V_{23}+V_{31}+
V_{45}+V_{56}+V_{64}\,,
\label{int6}
\eeq
Then we have, in analogy with (\ref{PPproj}):
\beq
P U \otimes D_6^{\infty} = P U_0 \otimes P D_6 = P U_0 \otimes D_6^{\infty}\,.
\eeq  
In ~\cite{JBMBGV} it has been shown that, on the rhs of (\ref{D6inteq}),
the $D_4^I$ pieces in the fourth and the fifth term contain  
transitions of a two-Pomeron state to a two-Odderon state which, at large 
$N_c$, are suppressed. As a result, in the large-$N_c$ limit
Eq.~(\ref{D6inteq}) takes the simple form 
\beq
S_6 D_6^{\infty} = P D_{60} + Z \otimes D_2 + P U_0 \otimes D_6^{\infty}\,,
\label{D6infinityEQ}
\eeq     
where $Z= lim_{N_c \to \infty} P \tilde{Z}$. This equation defines 
the diffractive $POO$ vertex, $Z$.

\begin{figure}
\begin{center}
\epsfig{file=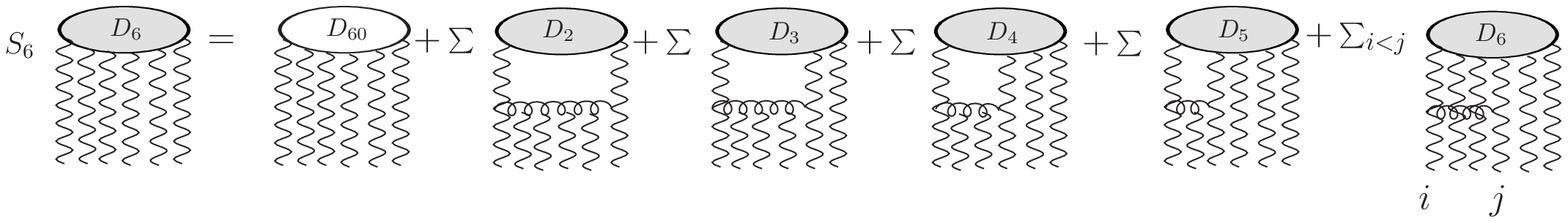,width=17cm,
}
\caption{ Equation for $D_6$.}
\end{center}
\label{figD6}
\end{figure}

The solution to this equation has a transparent physical meaning.
It is given by attaching two odderon Green functions (formed by 
gluons (123) and (456))  to the two inhomogeneous terms.
In the first term, the Odderons couple directly
to the $q\bar q$ pair (double Odderon exchange), in the second 
term the Odderons couple to the vertex $Z$ which describes diffractive 
gluon production.
After convoluting the Odderon Green's functions with appropriate
impact factors (e.g. couplings to the proton), one obtains cross sections 
for `diffraction with Odderon exchange'. 

To find the connection between the diffractive vertex $Z$ and the 
$POO$ vertex $W$ defined in ~\cite{BE}, we proceed in the 
same way as in the previous section. We start from the decomposition
of $D_6$ into symmetric and antisymmetric pieces, 
i.e. we invoke the reduction procedure which leads to
\beq
D_6 = D_6^R + D_6^I\,.
\label{D6decomposition}
\eeq
In fact, as discussed in ~\cite{BE}, $D_6^I$ is not yet fully
reduced and requires a further step of reduction. However, for our
purpose of discussing the splitting into two odderon configurations,
this is second step not necessary.
The non-reggeizing (``irreducible'') piece
$D_6^I$ satisfies the equation 
\beq
S_{6}D_6^I=W\otimes D_2+\sum L +\sum I +\sum J
+\sum K_{2 \to 4}\otimes D_4^I+\sum K_{2 \to 3}\otimes D_5^I+\sum U
\otimes D_6^I\,.
\label{D6Ieqfull}
\eeq
which defines the $POO$ vertex $W$. 
Projecting with $P$ onto the 2-Odderon state, and taking the 
large-$N_c$ limit, we use the results of ~\cite{JBMBGV}:
only the first and the last term on the rhs of (\ref{D6Ieqfull}) survive.
Substituting the relation $D_6^I=D_6-D_6^R$, we obtain
\beq
S_{6} P (D_6-D_6^R)=P W\otimes D_2+ P U \otimes ( D_6-D_6^R)\,,
\label{D6Ieqstep1}
\eeq
which leads to 
\beq
S_{6} D_6^\infty=(S_6-P U_0\otimes)D_6^R+ P W\otimes D_2+PU_0 \otimes
D_6^\infty\,.
\label{D6eq2}
\eeq
Comparison with eq. (\ref{D6infinityEQ}) gives the relations between
the two vertices:
\beq
P W\otimes D_2= P D_{60}+Z\otimes D_2-P(S_6-U_0\otimes)D_6^R
\label{WfromZ}
\eeq
and
\beq
Z\otimes D_2=P \left( W\otimes D_2 +(S_6-U_0\otimes)D_6^R- D_{60} \right)\,.
\label{ZfromW}
\eeq

\subsection{The explicit form of the diffractive POO vertex}
We now turn to the explicit form of $Z$. 
As shown in ~\cite{BE}, the explicit form of $D_6^R$ in the two-odderon
colour channel is given by
\[
D_6^R=C\Big\{
\sum_{i=1}^6D_2(i)-\sum_{i\neq k=1}^3D_2(ik)\]\beq
-\sum_{i\neq k=4}^6D_2(ik)-\sum_{i=1}^3\sum_{k=4}^6D_2(ik)+
\sum_{i\neq k=1}^3\sum_{l=4}^6D_2(ikl)+D_2(123)\Big\}\,,
\label{D6Rexpr}
\eeq
where the odderons are assumed to carry total momenta
$123=-456=q$. The coefficient $C$ already contains the color projection 
and has the form: 
\beq
C=g^4d^{abcdef}d_{abc}d_{def}=g^4\frac{(N_c^2-1)^2(N_c^2-4)^2}{8N_c^3}\,.
\label{C6}
\eeq
As in the previous section, in the following we will suppress this coefficient,
but keep in mind that, at the end, it has to be re-added in front of our 
final expressions.
The vertex $Z\otimes D_2$ can be viewed as an operator 
which acts on the two odderons. Each oddderon state is
symmetric in its three momentum variables, and the whole vertex is symmetric 
under the interchange of the two odderons.
So the quantity we have to study
\beq
X=(S_6-U_0)D_6^R-D_{60}
\label{X6ini}
\eeq
is to be convoluted with functions which are symmetric in the momenta (123)
and (456), and also symmetric under the interchange $(123)\leftrightarrow 456$.
Since the operator $S_6-U_0$ also possesses this symmetry,
it follows that all terms inside each of the sums in (\ref{D6Rexpr}) will give
equal contributions, that is, for our purpose we may simply write
\beq
D_6^R=6D_2(1)-6D_2(12)-9D_2(14)+9D_2(124)+D_2(123)\,.
\label{D6Rred}
\eeq

Different terms in the operator $(S_6-U_0)$ are again to be treated in a
different manner. The simplest term with the factor $j-1$
according to (\ref{BFKL}) leads to
\[
(j-1)D_6^R=D_{60}+6G(1,0,-1)-6G(12,0,-12)-9G(14,0,-14)\]\beq+
9G(124,0,-124)+G(123,0,-123)\,.
\label{parzjX6}
\eeq
Terms from $-\sum_{i=1}^6\o(i)$ will cancel with similar terms
coming from the remaining part of the operator. So, in fact, 
we have to study only 
$U_0 D_6^R$.  On defining
\beq
\tilde{X}=-2U_0 D_6^R\, ,
\label{parzexpr}
\eeq
the final $X$ will be given by
\beq
X=\frac{1}{2}\Big[2\Big(j-1-\sum_{i=1}^6\o(i)\Big)D_6^R+\tilde{X}\Big]
-D_{60}\,.
\label{X6}
\eeq

Using the identities (\ref{bootrel}) and (\ref{shiftrel}) one can express, 
in (\ref{parzexpr}), all terms via the trajectory function $\omega(i)$ 
and the function $W$. Subsequently one reexpresses
functions $W$ via the $G$'s and verifies that all terms proportional 
to $\omega$'s cancel.
The procedure closely follows the one for the triple pomeron vertex 
presented in the preceding section, but it requires considerably more algebra.
We present it in the Appendix 2.
Our final expression for $X$ in terms of $G's$ is
\[
X=9G(14,0,-14)-9G(124,0,-124)+G(123,0,-123)
-6G(12,3,-123)+6G(1,23,-123)\]\beq
-18G(4,1,-14)+18G(4,12,-124)+18G(12,4,-124)
-18G(14,23,56)\,.
\label{X6final}
\eeq
\subsection{The symmetry structure of $X$ and the final diffractive vertex}
The vertex $W$ in ~\cite{BE} has a peculiar symmetry structure 
~\cite{BLV,JBMBGV}.
It can be made explicit by introducing operators $\hat{S}$ and $\hat{P}$ 
which act on antisymmetric functions in the following way:
\beq
\hat{S}\phi(1,2)=\frac{1}{2}\sum_{(123)}[\phi(12,3)-\phi(3,12)]\,,
\eeq
where the sum is taken over the cyclic permutations of (123),
and
\beq
\hat{P}\phi(1,2)=\frac{1}{2}[\phi(123,0)-\phi(0,123)]\,.
\eeq
Then, ignoring the overall coefficient $C$ (Eq. (\ref{C6})),
the symmetric vertex $W$ can be cast into the form
\beq
W(1,2,3|4,5,6)=-4(\hat{S}_1-\hat{P}_1)f(1,2|4,5)
(\hat{S}^{\dagger}_2-\hat{P}^{\dagger}_2)\,.
\eeq
Here
\beq
f(1,2|4,5)=
\frac{1}{4}[G(1,24,5)-G(2,14,5)-G(1,25,4)+G(2,15,4)]\equiv {\cal A}G(1,24,5)\,.
\label{fsymm}
\eeq
The operator $\cal A$ antisymmetrizes in each pair of
variables 12 and 34, whereas the 
operators $\hat{S}_1$, $\hat{P}_1$ act on the variables 1 and 2,
and the operators $\hat{S}_2$, $\hat{P}_2$ on the variables 4 and 5.

One may ask if also the contribution $X$ to the vertex $Z$ derived
above can be cast  into a similar form,  with the function $f$ being replaced 
by a different function $g$ which again can be expressed in terms of $G$'s.
Obviously, apart from (\ref{fsymm}), we can have only two
antisymmetric functions
\beq
g^{(1)}={\cal A}G(14,2,5),\ \ {\rm and}\ \
g^{(2)}={\cal A}G(14,5,2)\,.
\eeq
Due to symmetry $(123)\leftrightarrow(456)$, after integration with 
two odderon Green's functions, they will give the same result. 
So it is sufficient to study
\beq
U_0^{(1)}=(\hat{S}_1-\hat{P}_1)g^{(1)}
(\hat{S}^{\dagger}_2-\hat{P}^{\dagger}_2)\,.
\eeq
Performing the necessary operations and using the symmetries we find
\[
4U_0^{(1)}=9G(4,1,-14)-9G(4,12,-124)-9G(12,4,124)+9G(14,23,56)-
3G(1,0,-1)\]\beq+3G(12,0,-12)+3G(12,3,-123)-3G(1,23,-123)-G(123,0,456).
\eeq
Comparing with our expression for $X$ we find
\begin{eqnarray}
X & =& -8U_0^{(1)}
-6G(1,0,-1)+6G(12,0,-12)+9G(14,0,-14)\nonumber \\
&&-9G(124,0,-124)-G(123,0,-123) \nonumber \\
& = & -8U_0^{(1)}+H_2D_6^R \,,
\end{eqnarray}
where $H_2(1,2)=-\o(1)-\o(2)- V_{12}$ is the BFKL
"Hamiltonian".
This allows to write the complete diffractive vertex $Z$ in the form
\[
Z(1,2,3|4,5,6)=\]\beq
-4 (\hat{S}_1-\hat{P}_1){\cal A}\Big(
G(1,24,5)+G(14,2,5)+G(14,5,2)\Big)
(\hat{S}^{\dagger}_2-\hat{P}^{\dagger}_2)+H_2 D_6^R(1,2,3|4,5,6)\,.
\eeq
Thus it almost matches the structure
of the symmetric vertex, except for the extra term, which has quite
a transparent meaning. Asymptotically, at high rapidities,
\beq
H_2D_6^R=-\Delta D_6^R\,,
\eeq
where $\Delta$ is the BFKL intercept.

The obtained diffractive vertex is infrared finite by itself
(and conformally invariant), since it is a sum of infrared finite and
conformally invariant functions $G$. One may ask if the infrared stability
is preserved when this vertex is coupled to the two odderons, which
involves an integration over all six momenta 1,2,..6 with a denominator
$\prod_{i=1}^6 k_i^2$.
The answer is affirmative, since, in fact, the odderons are better behaved
at $k_i^2\to 0$ than one might deduce from these denominators.
Let us take, as an example, the odderon constructed in ~\cite{BLV}. 
It has a structure
\[
O(1,2,3)=\frac{1}{k_1^2k_2^2k_3^2}\Big( P_a(12,3)+P_a(23,1)+P_a(31,2)\Big)\,,
\]
where $P_a(1,2)$ is an amputated antisymmetric pomeron state.
Consider the limit $k_1\to 0$. The bracket becomes 
\[ P_a(2,3)+P_a(23,0)+P_a(3,2),\]
which is zero, since $P_a$ is antisymmetric and goes to zero as one
of its arguments goes to zero. So the pole singularity at $k_1^2\to 0$
is softened.
As a result, the intergration of $k_1$ and of the other momenta
will not encounter any difficulties in the infrared region.

\section{Conclusions}
In this paper we have discussed a particular aspect of both the triple Pomeron
vertex and the Pomeron$\to$Odderon-Odderon vertex in pQCD. For the
calculation of diffractive cross sections, it is useful to
define vertices which differ from those which
appear in inclusive cross sections or in the context
of QCD reggeon field theory. We have named them `diffractive vertex'.
As to the triple Pomeron case, we have reviewed and analysed the relation
between this diffractive vertex and the standard triple Pomeron vertex.
This analysis has allowed to draw conclusions for the $s$-channel content
of the kernel of the BK equation. We then have applied the same line of
arguments to the Pomeron$\to$Odderon-Odderon vertex, and we have derived
the new diffractive vertex to be used in the computation of diffractive
states with Odderon exchange. For simplicity, we have restricted ourselves to
the large-$N_c$ limit.

\section{Acknowledgements}
M.A.B is grateful for the hospitality and financial support of the
Universities
of Hamburg and Bologna, where part of this work was done.
G.P.V. is very grateful to the Hamburg University for the hospitality.
This work has been supported by NATO Grant PST.CLG.980287 
\section{Appendix 1. Terms with $\omega$'s in the vertex PPP}
To check the cancellation of all the terms with $\omega$'s we now calculate
the last term which adds to the $G$'s in the expression (\ref{Xtildeexpr}).
We write those in $\tilde{X}$ as $Y$, and in particular $Y_a$ is
related to the $\tilde{X}_a$ term, where $a=1,2,3,4,12,13,14$
is any string of the momenta present in the $D_4^R$ expression.

Start with the term $\tilde{X}_1$:
\beq
Y_1=D_2(1)[-2\o(34)-2\o(3)-2\o(4)+\o(2)+\o(1)]
+D_2(12)[\o(2)-\o(12)]\, .
\eeq
An analogous result is obtained for the other the cases $\tilde{X}_2$,
$\tilde{X}_3$ and $\tilde{X}_4$.
From $\tilde{X}_{12}$:
\beq
Y_{12}=D_2(12)[2\sum_{i=1}^4\omega(i)-2\o(12)-2\o(34)]\,.
\eeq

There are quite a few terms from $\tilde{X}_{13}$.
They sum into
\[
Y_{13}=D_2(13)\sum_{i=1}^4\o(i)
+D_2(1)[\o(3)-\o(34)]+D_2(2)[\o(4)-\o(34)]\]\beq
+D_2(3)[\o(1)-\o(12)]+D_2(4)[\o(2)-\o(12)]\, .
\eeq
A similar result is obtained for $Y_{14}$.
Now we consider the expression
\beq
Y=\sum_{i=1}^4 Y_i-Y_{12}-Y_{13}-Y_{14}=\sum_a c_a D_2(a)\, .
\eeq
The coefficient of the $D_2(1)$ term in $Y$ is given by
\beq
c_1=-2\o(34)+2\o(3)+2\o(4)+\o(2)+\o(1)-\o(4)+\o(34)-\o(3)+\o(34)=
\sum_{i=1}^4 \o(i) \,
\eeq
and similarly one gets $c_2=c_3=c_4=c_1$.
We consider now $c_{12}$:
\bea
c_{12}&=&2\o(12)+2\o(34)-2\sum_{i=1}^4\o(i)
+\o(2)-\o(12)+\o(1)-\o(12)\nonumber \\
&&+\o(4)-\o(34)+\o(3)-\o(34)=-\sum_{i=1}^4 \o(i) \,.
\eea
Finally let us consider $c_{13}$, which is immediately seen to be:
\beq
c_{13}=-\sum_{i=1}^4 \o(i)\,,
\eeq
and the same result holds for $c_{14}$.
This means we have proved that
\beq
Y=\sum_{i=1}^4 \o(i) D_4^R \, ,
\eeq
which is the desired result which allows the cancellation in $X$.
\section{Appendix 2. Useful formulas for the POO diffractive vertex}
\subsection{Action of $U_0$ on different terms in $D_6^R$}
We begin with
\beq
\tilde{X}_1=-2U_0D_2(1)=-(V_{12}+V_{23}+V_{31}+
V_{45}+V_{56}+V_{64})D_2(1).
\eeq
Terms independent of $1$ will give $\omega$'s:
\[
-(V_{23}+V_{45}+V_{56}+V_{64})D_2(1)\]\beq=
-2[\omega(23)-\omega(2)-\omega(3)+\omega(45)-\omega(4)-\omega(5)+
\omega(56)-\omega(5)-\omega(6)+\omega(64)-\omega(6)-\omega(4)].
\eeq
We are left with terms which contain
\beq
(V_{12}+V_{13})D_2(1)=
W(1,2,-12)+W(1,3,-13)-2W(1,0,-1).
\eeq
Summing all terms we get
\[
\tilde{X}_1=[2W(1,0,-1)-W(1,2,-12)-W(1,3,-13)]\]\[
-2[\omega(23)-\omega(2)-\omega(3)+\omega(45)-\omega(4)-\omega(5)\]\[+
\omega(56)-\omega(5)-\omega(6)+\omega(64)-\omega(6)-\omega(4)]D_2(1)
\]\[
=[2W(1,0,-1)-W(1,2,-12)-W(1,3,-13)]+\]\beq
[2\sum_{i=2}^3\omega(i)+4\sum_{i=4}^6\omega(i)-2\omega(23)-2\omega(45)-
2\omega(56)-2\omega(64)]D_2(1).
\eeq

Now we pass to the second term in (\ref{D6Rred}):
\beq
\tilde{X}_{12}=-2U D_2(12)=-(V_{12}+V_{23}+V_{31}+
V_{45}+V_{56}+V_{64})D_2(12).
\eeq
Terms which give $\omega$'s are:
\[
-(V_{12}+V_{45}+V_{56}+V_{64})D_2(12)\]\beq=
-2[\omega(12)-\omega(1)-\omega(2)+\omega(45)-\omega(4)-\omega(5)+
\omega(56)-\omega(5)-\omega(6)+\omega(64)-\omega(6)-\omega(4)].
\eeq
We are left with two terms which contain
\beq
V_{13}D_2(12)=W(-12,1,2)+W(-123,3,12)-W(2,13,-123)-
W(12,0,-12)
\eeq
and
\beq
V_{23}D(12)=W(-12,2,1)+W(-123,3,12)-W(1,23,-123)-
W(12,0,-12).
\eeq

Summing all the terms we get
\[
\tilde{X}_{12}=
-[W(-12,1,2)+W(-123,3,12)-W(2,13,-123)\]\[-
W(12,0,-12)
+W(-12,2,1)+W(-123,3,12)-W(1,23,-123)-
W(12,0,-12)]
\]\[
-2[\omega(12)-\omega(1)-\omega(2)+\omega(45)-\omega(4)-\omega(5)\]\[+
\omega(56)-\omega(5)-\omega(6)+\omega(64)-\omega(6)-\omega(4)]D_2(12)=
\]\[
=+[2W(12,0,-12)
-W(-12,1,2)\]\[-2W(-123,3,12)+W(2,13,-123)
-W(-12,2,1)+W(1,23,-123)]\]\beq
+[2\omega(1)+2\omega(2)+4\sum_{i=4}^6\omega(i)-2\o(12)
-2\omega(45)-2\omega(56)-2\omega(64)]D_2(12).
\eeq

From the third term in (\ref{D6Rred}) we have
\beq
\tilde{X}_{14}=-(V_{12}+V_{23}+V_{31}+
V_{45}+V_{56}+V_{64})D_2(14).
\eeq

Terms which give $\omega$'s are:
\beq
-(V_{23}+V_{56})D_2(12)=
-2[\omega(23)-\omega(2)-\omega(3)+
\omega(56)-\omega(5)-\omega(6)]D_2(14).
\eeq
Now we have 4 non-trivial terms containing
\beq
V_{12}D_2(14)=W(-14,1,4)+W(-124 ,2,14)-W(4,12,-124)-W(14,0,-14),
\eeq
\beq
V_{13}D(14)=W(-14,1,4)+W(-134 ,3,14)-W(4,13,-134)-W(14,0,-14),
\eeq
\beq
V_{45}D(14)=W(-14,4,1)+W(-145 ,5,14)-W(1,45,-145)-W(14,0,-14),
\eeq
\beq
V_{46}D(14)=W(-14,4,1)+W(-146 ,6,14)-W(1,46,-146)-W(14,0,-14).
\eeq
Summing all:
\[
\tilde{X}_{14}=[
-W(-14,1,4)-W(-124 ,2,14)+W(4,12,-124)\]\[+W(14,0,-14)
-W(-14,1,4)-W(-134 ,3,14)+W(4,13,-134)+W(14,0,-14)\]\[
-W(-14,4,1)-W(-145 ,5,14)+W(1,45,-145)+W(14,0,-14)\]\[
-W(-14,4,1)-W(-146 ,6,14)+W(1,46,-146)+W(14,0,-14)]\]\[
-2[\omega(23)-\omega(2)-\omega(3)+
\omega(56)-\omega(5)-\omega(6)]D_2(14)\]\[=
[4W(14,0,-14)
-2W(-14,1,4)-W(-124 ,2,14)+W(4,12,-124)\]\[
-W(-134 ,3,14)+W(4,13,-134)\]\[
-2W(-14,4,1)-W(-145 ,5,14)+W(1,45,-145)
-W(-146 ,6,14)+W(1,46,-146)]\]\beq+
[2\omega(2)+2\omega(3)+2\omega(5)+2\omega(6)
-2\omega(23)-2\omega(56)]D_2(14).
\eeq

The fourth term gives contributions like
\beq
\tilde{X}_{124}=-2U D_2(1)=(V_{12}+V_{23}+V_{31}+
V_{45}+V_{56}+V_{64})D_2(124).
\eeq
Terms which give $\omega$'s are:
\beq
-(V_{12}+V_{56})D_2(124)=
-2[\omega(12)-\omega(1)-\omega(2)+
\omega(56)-\omega(5)-\omega(6)]D_2(124)
\eeq
and again we have 4 nontrivial terms
\beq
V_{13}D(124)=W(-124,1,24)+W(-1234 ,3,124)-
W(24,13,-1234)-W(124,0,-124),
\eeq
\beq
V_{23}D(124)=W(-124,2,14)+W(-1234 ,3,124)-
W(14,23,-1234)-W(124,0,-124),
\eeq
\beq
V_{45}D(124)=W(-124,4,12)+W(-1245 ,5,124)
-W(12,45,-1245)-W(124,0,-124),
\eeq
\beq
V_{46}D(124)=W(-124,4,12)+W(-1246 ,6,124)
-W(12,46,-1246)-W(124,0,-124).
\eeq
We sum all to obtain
\[
\tilde{X}_{124}=[
-W(-124,1,24)-W(-1234 ,3,124)+
W(24,13,-1234)\]\[+W(124,0,-124)
-W(-124,2,14)-W(-1234 ,3,124)\]\[+
W(14,23,-1234)+W(124,0,-124)
-W(-124,4,12)-W(-1245 ,5,124)\]\[
+W(12,45,-1245)\]\[+W(124,0,-124)
-W(-124,4,12)-W(-1246 ,6,124)
+W(12,46,-1246)+W(124,0,-124)]\]\[
-2[\omega(12)-\omega(1)-\omega(2)+
\omega(56)-\omega(5)-\omega(6)]D_2(124)\]\[
=[4W(124,0,-124)
-W(-124,1,24)-2W(-1234 ,3,124)+W(24,13,-1234)\]\[
-W(-124,2,14)+W(14,23,-1234)
-2W(-124,4,12)\]\[-W(-1245 ,5,124)+W(12,45,-1245)
-W(-1246 ,6,124)+W(12,46,-1246)]\]\beq+
[2\omega(1)+2\omega(2)+2\omega(5)+2\omega(6)
-2\omega(12)-2\omega(56)]D_2(124).
\eeq
Finally the last term
\beq
\tilde{X}_{123}=-2U D_2(123)=(V_{12}+V_{23}+V_{31}+
V_{45}+V_{56}+V_{64})D_2(123).
\eeq
Here
all terms give $\omega$'s:
\[
-(V(12)+V(23)+V(31)+V(45)+V_{56}+V(64))D_2(124)\]\beq=
-2[\omega(12)+\omega(23)+\omega(13)+\omega(45)+\omega(56)+
\omega(46)-2\sum_{i=1}^6\omega(i)]D_2(123).
\eeq
So we find
\[
\tilde{X}_{123}=\]\beq
[4\sum_{i=1}^6\omega(i)-2\omega(12)-2\omega(23)-
2\omega(31)-2\omega(45)-2\omega(56)-2\omega(46)]D_2(123).
\eeq

Our next step is to transform these expressions to functions
$G$ using (\ref{WGrel}).
This changes $g^2N_cW$ into $-G$ and
gives additional terms with $\omega$'s. It is expected that these
additional terms with $\omega$'s plus already existing in our expressions
will cancel leaving the final expression for the vertex entirely in terms
of $G$.  Calculation shows that this is indeed so (see below).

As a result we  are left with terms with $G's$. Summed as indicated in
(\ref{D6Rred}) they give a long expression
\[
\tilde{X}=
-12G(1,0,-1)+6G(1,2,-12)+6G(1,3,-13)
+12G(12,0,-12)\]\[
-6G(-12,1,2)-12G(-123,3,12)+6G(2,13,-123)
-6G(-12,2,1)\]\[+6G(1,23,-123)
+36G(14,0,-14)\]\[
-18G(-14,1,4)-9G(-124 ,2,14)+9G(4,12,-124)
-9G(-134 ,3,14)+9G(4,13,-134)\]\[
-18G(-14,4,1)-9G(-145 ,5,14)+9G(1,45,-145)
-9G(-146 ,6,14)+9G(1,46,-146)\]\[
-36G(124,0,-124)
+9G(-124,1,24)+18G(-1234 ,3,124)\]\[-9G(24,13,-1234)
+9G(-124,2,14)-9G(14,23,-1234)\]\[
+18G(-124,4,12)+9G(-1245 ,5,124)-9G(12,45,-1245)\]\beq
+9G(-1246 ,6,124)-9G(12,46,-1246)+ [\omega \,\,\, {\rm terms}].
\eeq
It can be considerably simplified by use of symmetry properties.
Finally one finds, taking in to account the $\omega$ terms,
as shown in the next subsections, the following result
\[
\tilde{X}=-2UD_6^R=-12G(1,0,-1)+12G(12,0,-12)+36G(14,0,-14)
-36G(124,0,-124)
\]\[-12G(12,3,-123)+12G(1,23,-123)
-36(4,1,-14)
+36G(4,12,-124)\]\beq+36G(12,4,-124)
-36G(14,23,56)+2\sum_{i=1}^6\o(i)D_6^R\, .
\eeq
According to (\ref{X6}) to find $X$ we have to add to this expression
twice the term (\ref{parzjX6}),
take one half of the sum and subtract $D_{60}$. This gives our final
expression (\ref{X6final}) for $X$
in terms of $G$'s.

\subsection{Terms with $\omega$'s}
To check the cancellation of all the terms with $\omega$'s  we have to
write out all of them.
We write those in $\tilde{X}$ as $Y$ and we want to see the total
amount of them in $X$
\beq
Y-2\sum_{i=1}^6\o(i)D_6^R
\eeq
as given by eq. (50). Then our aim is to see if all terms cancel.

Start with $\tilde{X}_1$. It has the following $W$'s:
\[ 2W(1,0,-1)\to -2D_2(1)[\o(0)-\o(1)]-2D_2(1)[\o(0)-\o(1)],\]
\[-W(1,2,-12)\to D_2(1)[\o(2)-\o(1)]+D_2(12)[\o(2)-\o(12)],\]
\[-W(1,3,-13)\to D_2(1)[\o(3)-\o(1)]+D_2(13)[\o(3)-\o(13)].\]
Adding these new terms to the old ones we have a term with $D_2(1)$:
\[D_2(1)[2\sum_{i=2}^3\omega(i)+4\sum_{i=4}^6\omega(i)-2\omega(23)\]\[
-2\omega(45)-
2\omega(56)-2\omega(64)-4\o(0)+4\o(1)+\o(2)+\o(3)-\o(1)-\o(1)],\]
which together with others and subtracting $2\sum_{i=1}^6\o(i)D_2(1)$
gives
\[
Y_1=D_2(1)[\o(2)+\o(3)+2\sum_{i=4}^6\omega(i)
-2\omega(23)-2\omega(45)\]\beq-
2\omega(56)-2\omega(64)]+D_2(12)[\o(2)-\o(12)]+D_2(13)[\o(3)-\o(13)].
\eeq
Here we used $\o(0)=0$, thanks to the good behaviour due to I.R.
cancellations. 

Now $W$'s from $\tilde{X}_{12}$:
\[2W(12,0,-12)\to -2D_2(12)[\o(0)-\o(12)] -2D_2(12)[\o(0)-\o(12)],\]
\[-W(-12,1,2)\to D_2(12)[\o(1)-\o(12)]+D_2(2)[\o(1)-\o(2)],\]
\[-2W(-123,3,12)\to 2D_2(123)[\o(3)-\o(123)]+2D_2(12)[\o(3)-\o(12)],\]
\[+W(2,13,-123)\to -D_2(2)[\o(13)-\o(2)]-D_2(123)[\o(13)-\o(123)],\]
\[-W(-12,2,1)\to D_2(12)[\o(2)-\o(12)]+D_2(1)[\o(2)-\o(1)],\]
\[W(1,23,-123)\to- D_2(1)[\o(23)-\o(1)]-D_2(123)[\o(23)-\o(123)].\]
Term with $D(12)$ results
\[
D_2(12)[2\omega(1)+2\omega(2)+4\sum_{i=4}^6\omega(i)-2\omega(12)-
2\omega(45)-2\omega(56)\]\[-2\omega(64)-4\o(0)+4\o(12)+\o(1)-\o(12)
+\o(2)-\o(12)+2\o(3)-2\o(12)].\]
Summing all and subtracting $2\sum_{i=1}^6\o(i)D_2(12)$ we obtain
\[
D_2(12)[\omega(1)+\omega(2)+2\sum_{i=4}^6\omega(i)
-2\o(12)-2\omega(45)-2\omega(56)-2\omega(64)]\]\beq
+D_2(1)[\o(2)-\o(23)]+D_2(2)[\o(1)-\o(13)]
+D_2(123)[2\o(3)-\o(13)-\o(23)].\eeq

Terms from $\tilde{X}_{14}$ The $W$'s give
\[4W(14,0,-14)\to-4D_2(14)[\o(0)-\o(14)]-4D_2(14)[\o(0)-\o(14)],\]
\[-2W(-14,1,4)\to+2D_2(14)[\o(1)-\o(14)]+2D_2(4)(\o(1)-\o(4)],\]
\[-W(-124 ,2,14)\to+D_2(124)[\o(2)-\o(124)]+D_2(14)[\o(2)-\o(14)],\]
\[+W(4,12,-124)\to -D_2(4)[\o(12)-\o(4)]-D_2(124)[\o(12)-\o(124)],\]
\[-W(-134 ,3,14)\to +D_2(134)[\o(3)-\o(134)]+D_2(14)[\o(3)-\o(14)],\]
\[+W(4,13,-134)\to-D_2(4)[\o(13)-\o(4)]-D_2(134)[\o(13)-\o(134)],\]
\[-2W(-14,4,1)\to+2D_2(14)[\o(4)-\o(14)]+2D_2(1)[\o(4)-\o(1)],\]
\[-W(-145 ,5,14)\to+D_2(145)[\o(5)-\o(145)]+D_2(14)[\o(5)-\o(14)],\]
\[+W(1,45,-145)\to-D_2(1)[\o(45)-\o(1)]-D_2(145)[\o(45)-\o(145)],\]
\[-W(-146 ,6,14)\to+D_2(146)[\o(6)-\o(146)]+D_2(14)[\o(6)-\o(14)],\]
\[+W(1,46,-146)\to-D_2(1)[\o(46)-\o(1)]-D_2(146)[\o(46)-\o(146)].\]
Term with $D_2(14)$ results
\[
D_2(14)[2\omega(2)+2\omega(3)+2\omega(5)+2\omega(6)
-2\omega(23)-2\omega(56)\]\[
-8\o(0)+8\o(14)+2\o(1)-2\o(14)+\o(2)\]\[-\o(14)+\o(3)-\o(14)+2\o(4)
-2\o(14)+\o(5)-\o(14)+\o(6)-\o(14)].\]
Summing all and subtracting $2\sum_{i=1}^6\o(i)D_2(14)$
gives
\[
D_2(14)[\o(2)+\o(3)+\o(5)+\o(6)\]\[-2\omega(23)-2\omega(56)]
+D_2(1)[2\o(4)-\o(45)-\o(46)]+D_2(4)[2\o(1)-\o(12)-\o(13)]\]\[
+D_2(124)[\o(2)-\o(12)]+D_2(134)[\o(3)-\o(13)]\]\beq
+D_2(145)[\o(5)-\o(45)]+D_2(146)[\o(6)-\o(46)].\eeq

Terms from $\tilde{X}_{124}$. From $W$'s we get
\[4W(124,0,-124)\to -4D_2(124)[\o(0)-\o(124)]-4D_2(124)[\o(0)-\o(124)],\]
\[-W(-124,1,24)\to D_2(124)[\o(1)-\o(124)]+D_2(24)[\o(1)-\o(24)],\]
\[-2W(-1234 ,3,124\to 2D_2(56)[\o(3)-\o(56)]+2D_2(124)[\o(3)-\o(124)],\]
\[+W(24,13,-1234)\to -D_2(24)[\o(13)-\o(24)]-D_2(56)[\o(13)-\o(56)],\]
\[-W(-124,2,14)\to D_2(124)[\o(2)-\o(124)]+D_2(14)[\o(2)-\o(14)],\]
\[+W(14,23,-1234)\to -D_2(14)[\o(23)-\o(14)]-D_2(56)[\o(23)-\o(56)],\]
\[-2W(-124,4,12)\to 2D_2(124)[\o(4)-\o(124)]+2D_2(12)[\o(4)-\o(12)],\]
\[-W(-1245 ,5,124)\to D_2(36)[\o(5)-\o(36)]+D_2(124)[\o(5)-\o(124)],\]
\[+W(12,45,-1245)\to -D_2(12)[\o(45)-\o(12)]-D_2(36)[\o(45)-\o(36)],\]
\[-W(-1246 ,6,124)\to D_2(35)[\o(6)-\o(35)]+D_2(124)[\o(6)-\o(124)],\]
\[+W(12,46,-1246)\to D_2(12)[\o(46)-\o(12)]-D_2(35)[\o(46)-\o(35)].\]
Term with $D_2(124)$ results
\[
D_2(124)[2\omega(1)+2\omega(2)+2\omega(5)+2\omega(6)
-2\omega(12)-2\omega(56)\]\[-8\o(0)+8\o(124)+\o(1)-\o(124)
+2\o(3)-2\o(124)+\o(2)-\]\[\o(124)+2\o(4)-2\o(124)+\o(5)-\o(124)
+\o(6)-\o(124)].\]
Summing all and subtracting $2\sum_{i=1}^6\o(i)D_2(124)$
\[
D_2(124)[\o(1)+\o(2)+\o(5)+\o(6)-2\omega(12)-2\omega(56)]\]\[
+D_2(12)[2\o(4)-\o(45)-\o(46)]+D_2(14)[\o(2)-\o(23)]\]\[
+D_2(24)[\o(1)-\o(13)]+D_2(35)[\o(6)-\o(46)]\]\beq
+D_2(36)[\o(5)-\o(45)]+D_2(56)[2\o(3)-\o(13)-\o(23)].\eeq

From the las term $\tilde{X}_{123}$ we trivially find
\beq
Y_{123}=D_2(123)[2\sum_{i=1}^6\o(i)-2\o(12)-2\o(23)-2\o(31)-2\o(45)
-2\o(56)-2\o(46)].
\eeq

To see that all these $\omega$ terms cancel we have to take
into account the symmetries and
study the sum
\beq
6Y_1-6Y_{12}-9Y_{14}+9Y_{124}+Y_{123}.
\eeq
Summing all the $Y$'s and making the necessary permutations we get
a sum of five different structures
\[ a_1D_2(1)+a_{12}D_2(12)+a_{14}D_2(14)+a_{124}D_2(124)+a_{123}D_2(123).\]
It is then straightforward to see that all the coefficients
$a_1$, $a_{12}$, $a_{14}$ $a_{124}$ and $a_{123}$ are equal to zero.
So indeed all $\omega$ terms in $X$ cancel.

\end{document}